\documentclass[aps,prb,twocolumn,floatfix,superscriptaddress,llncs]{revtex4-1}  
\usepackage{graphicx,epsfig,dcolumn,amssymb,amsmath}
\usepackage{booktabs}
\usepackage{float}
\usepackage{color}
\usepackage{lipsum}
\usepackage{hyperref}

\usepackage{xcolor}

\makeatletter
\newcommand{\printfnsymbol}[1]{%
  \textsuperscript{\@fnsymbol{2}}%
}

\begin{document}

\title{Edge and Width Dependent Electronic Properties of Nanoribbons of Manganese Oxide}

\author{Yigit Sozen}
\affiliation{Department of Photonics, Izmir Institute of Technology, 35430, 
Izmir, Turkey} 

\author{Ugur C. Topkiran}
\affiliation{Department of Physics, Izmir Institute of Technology, 35430, 
Izmir, Turkey}

\author{Hasan Sahin}
\affiliation{Department of Photonics, Izmir Institute of Technology, 35430, 
Izmir, Turkey}

\begin{abstract}

In the present work, the structural, magnetic, and electronic properties of the two- and one-dimensional honeycomb structures of recently synthesized MnO [Zhang et al. Nat. Commun., \textbf{20}, 1073-1078 (2021)] are investigated by using first principles calculations. Our calculations show that the single layer 2D MnO crystal has a degenerate antiferromagnetic (AFM) ground state and a relatively less favorable ferromagnetic (FM) state. In addition, magnetic anisotropy calculations unveil that the easy-axis direction for magnetism originating from unpaired electron states in manganese atoms is normal to the crystal plane. Electronically, while the FM-MnO is a direct semiconductor with a narrow bandgap, AFM phases display large indirect bandgap semiconducting behavior. Moreover, calculations on nanoribbons of MnO reveal that zigzag edged ribbons display metallic bahavior, whereas armchair edged nanoribbons are semiconductors. Magnetically, for both zigzag- or armchair-edged nanoribbons, AFM order perpendicular to the nanoribbon growth direction is found to be favorable over the other AFM and FM orders.  
Moreover, depending on the edge symmetry and ribbon width, forbidden band gap values of nanoribbons display distinct family behaviors.

\end{abstract}

\maketitle

\section{Introduction}
Over the past decade, it has been shown that the dimensional reduction not only allows scaling of the device sizes but also brings outstanding physical phenomena due to the quantized motion of electrons and holes, thus paving the way for the development of compact devices with novel features \cite{liu2017emerging}. In 2004, nanotechnology has started to conduct beyond the nanoscale with the realization of atomically thin derivative of two-dimensional (2D) graphite sheet, graphene, which displays unconventional properties \cite{novoselov2004electric}. The unique properties of graphene mainly stems from the dominant quantum effects in atomic scale that brings enhanced electronic and thermal transport \cite{geim2010rise,neto2009electronic}. Graphene-triggered research have also revealed the existence of brand new 2D materials such as silicene \cite{vogt2012silicene, cahangirov2009two}, germanene\cite{cahangirov2009two}, hexagonal BN \cite{li2016atomically}, transition metal dichalcogenides (TMDs) \cite{chhowalla2013chemistry,mak2010atomically,wang2012electronics,ross2014electrically,georgiou2013vertical,radisavljevic2011single,tongay2014monolayer} and phosphorene \cite{liu2014phosphorene}. Although these materials are being studied extensively by researchers, the number of ultrathin materials exhibiting magnetic properties is very few. Among the ultra-thin 2D crystal materials that have been discovered so far, especially CrI$_3$ \cite{mcguire2015coupling}, Cr$_2$Ge$_2$Te$_6$\cite{xing2017electric}, Fe$_2$GeTe$_2$ \cite{deng2018gate}, FePS$_3$ \cite{wang2016raman}, FeCl$_2$ \cite{zhou2020atomically}, VS$_2$ \cite{ma2012evidence}, VSe$_2$ \cite{ma2012evidence},  VI$_3$ \cite{he2016unusual}, VCl$_3$ \cite{he2016unusual}, MnSe$_2$ \cite{o2018room}, and RuCl$_3$ \cite{banerjee2016proximate} come to the fore.

Further in plane dimensional reduction of 2D crystal structures led to the formation of nanoribbons (NRs) in which the propagation of charge carriers are confined to one dimension (1D) along the ribbon growth direction \cite{yagmurcukardes2016nanoribbons, son2006energy, lee2005magnetic}. The large-scale fabrication of 1D derivatives of variety range of 2D materials, such as graphene  \cite{barone2006electronic,cai2010atomically,vo2014large,jacobberger2015direct}, MoS$_2$ \cite{mak2016photonics, chhowalla2013chemistry}, and phosphorene \cite{masih2016controlled}, have been reported. These advances promoted their integration into device applications of spintronics \cite{ruffieux2016surface, kong2014spin}, optoelectronics \cite{blankenburg2012intraribbon}, and quantum information technologies \cite{guo2009quantum, chen2015theoretical, tan2020highly}. 

Among the 1D family, graphene NRs have been most intensively studied prospects, which are useful for future nanoelectronics \cite{wang2021graphene}. These materials are simply classified according to their edge geometries, and each brings distinctive electronic and magnetic features. Resulting from the hexagonal symmetry, zigzag and armchair edge formation is possible in graphene NRs. While the edge states induces metallicity in zigzag type structure, armchair edge termination opens a band gap, which is controllable changing the ribbon width \cite{lee2005magnetic,son2006energy}. Moreover, the passivation edge states with H atoms in zigzag NR leads to direct band gap formation. Unlike the nonmagnetic nature of armchair NR, intrinsic ferromagnetic order in opposite edge states of zigzag NR were reported to couple antiferromagnetically with each other \cite{lee2005magnetic}.

Although nanoribbons of various materials have been extensively studied, it has not been investigated how the properties of a 2D material that exhibits intrinsically magnetism changes with further confinement to 1D. As demonstrated by a recent experimental study,  single layer easy synthesis of graphene-like 2D crystalline structures of TiO$_2$, Fe$_2$O$_3$, CoO, Ni$_2$O$_3$, MnO and Cu$_2$O can be achieved by performing a strictly controlled oxidation at the metal-gas interfaces \cite{zhang2021hexagonal}. Among these materials, which have been experimentally synthesized, manganese oxide (MnO), which offers a rich playground in terms of its magnetic properties, was chosen as the target material of our study. Here, we carry out first-principles calculations in order to investigate the structural, electronic, and magnetic properties of hexagonal MnO in one and two dimensions. 

The paper is organized as follows: in Sec. II we provide the details of our computational methodology. Magnetic and electronic properties of 2D MnO are discussed in Sec. III(A). In addition, Sec. III(B) presents the edge- and width-dependent properties of NRs.

\section{Computational Methodology}\label{sec:computational}

Structural relaxation, electronic dispersion and magnetic ground state calculations were carried out on the basis of density functional theory (DFT) implemented in the Vienna ab-initio Simulation Package (VASP)\cite{kresse1993ab, kresse1996efficient} based on plane wave projector augmented wave (PAW) \cite{kresse1999ultrasoft, blochl1994projector} pseudopotential formalism of spin-polarized density functional theory. Generalized gradient approximation (GGA) in the form of Perdew-Burke-Ernzerhof (PBE) \cite{perdew1996generalized} functional was used for exchange-correlation energy. The van der Waals (vdW) forces were taken into account using the DFT-D2 method of Grimme \cite{grimme2006semiempirical}. In addition, for determination of the eaxy axis of the magnetic moments inside crystal structure non-collinear calculations were perfomed. Bader charge technique was utilized to obtain the final charge density on atoms constituting the crystal structure \cite{henkelman2006fast}.

To include the correlations between $d$-orbitals of Mn atoms, the difference between the on-site Coulomb parameter (U) and the exchange parameter (J), U$_{eff}$ = U-J, was taken to be 3.9 eV on the basis of DFT+U method \cite{dudarev1998electron}. The kinetic energy cutoff and the convergence criterion for the total energy was set to 650 eV and 10$^{-5}$ eV, respectively. 

A suffieciently large vaccum space of 18 \AA{} is inserted in order to avoid adjacent layer-layer interactions along the z-axis. For the structural optimization of the primitive unit cell of 2D and NR MnO, a 6$\times$5$\times$1 and 6$\times$1$\times$1 $\Gamma$-centered k-point meshes were used respectively, and they were doubled for accurate density of states calculations. Pressures on the unit cell were decreased to a value within $\pm$ 1.0 kBar in all three directions to achieve a fully relaxed structure. 

For various magnetic phases of 2D crystal, the cohe­sive energy per atom was calculated using the equation $E_{Coh} = \frac{1}{n_{tot}}[{n_{Mn}E_{Mn}+n_{O}E_{O}-E_{sys}}]$,  where $n_{Mn}$ is the number of Mn atoms, $n_{O}$ is the number of O atoms, $n_{tot}$ is the total number of atoms, $E_{Mn}$  is the energy of a single Mn atom, $E_{O}$ is the energy of a single O atom and $E_{sys}$ is the total energy of the system.

\section{Results and discussion}\label{sec:results}

\subsection{2D Single Layers of Manganese Oxide}

\begin{figure}[t]
\centering
  \includegraphics[width=8.5cm]{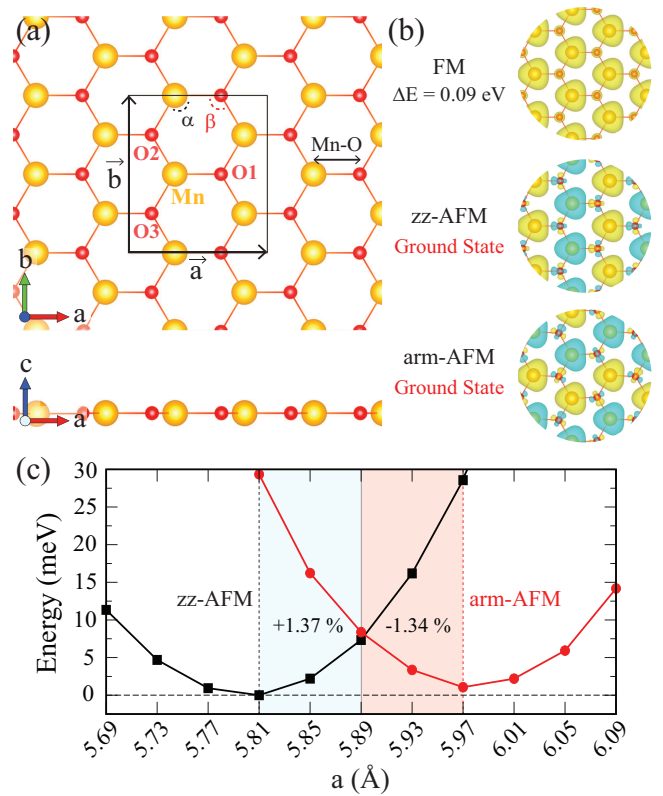}
  \caption{(a) Top and side views of the geometric structure of 2D MnO. Orange and red atoms represent Mn and O, respectively. (b) Spin density plots of FM, zz-AFM, and arm-AFM orders, at the isosurface value of $5 \times 10^{-3}$ e/\AA{}$^{3}$. $\Delta$E corresponds to energy difference per formula unit with respect to the ground state structure. (c) Variation in the ground state energy of zz-AFM and arm-AFM with respect to the applied strain along the armchair direction.}
  \label{fig:1}
\end{figure}

As shown in Fig. \ref{fig:1}a, 2D MnO displays graphene-like planar hexagonal lattice structure, which contains single Mn-O pair in its primitive unit cell. Presence of unpaired electrons on each Mn requires investigation of possible magnetic orders for determination of the ground state of the material. Therefore, the repeating unit is considered as a rectangular unit cell, to simulate ferromagnetic (FM), zigzag antiferromagnetic (zz-AFM) and armchair antiferromagnetic (arm-AFM) orders in the crystal lattice (see Fig. \ref{fig:1}b). Total energy calculations reveal that, both AFM configurations are energetically identical and more favorable than FM with the energy difference ($\Delta$E) of 0.09 eV per formula unit. 3D magnetic charge densities represented in Fig. \ref{fig:1}b indicate that AFM coupling inside the lattice is stabilized through the superexchange interactions between magnetic ions mediated by O atoms. Furthermore, it is found that each Mn atom induces a magnetic moment of 5 $\mu_B$ originating from the unpaired electrons of \textit{3d} orbitals.

\medskip
\def\arraystretch{1.4}
\begin{table*}[t]
\centering
\small
\caption{For the different magnetic phases of MnO, the optimized lattice parameters for rectangular unit cell, $a$ and $b$; atomic distances between Mn atom and neighboring O atoms, $d_{Mn-O_{1-3}}$; the internal angles within a single MnO hexagon, $\alpha$ and $\beta$; the charge transferred from a Mn atom to surrounding O atoms, ($\Delta \rho$); the magnetic moment per Mn atom, ($\mu$); the cohesive energy, $E_{Coh}$; the workfunction, $\Phi$; the band gap value $E_{gap}$, with the band gap type, D or I, corresponding to direct or indirect band gaps. }
  \label{table1}
  \begin{tabular*}{\textwidth}[t]{@{\extracolsep{\fill}}lccccccccccccc}
    \hline
    \hline
    &   &  $a$  & $b$  & $d_{Mn-O_{1}}$ & $d_{Mn-O_{2}}$  & $d_{Mn-O_{3}}$  &$\alpha$  &$\beta$  & $ \Delta \rho $  &  $\mu_{Mn}$  & $E_{Coh}$ & $\Phi$ & $E_{gap} $  \\
    &   & (\AA)  & (\AA)  & (\AA)  & (\AA)  & (\AA)  &  (deg)  &  (deg)  & ($e^-$)  &  ($\mu_{B}$)  & (eV)  & (eV)  & (eV)  \\
    \hline
    
    &  FM               & 5.97  &  6.89  & 1.99  & 1.99  & 1.99  & 120    &   120  & 1.3  & 5  & 4.48  & 3.03  & 0.17 (D)  \\
    &  zz-AFM   & 5.81  &  6.97  & 1.97  & 1.98  & 1.98  & 118.2  & 123.5  & 1.3  & 0  & 4.52  & 3.72  & 1.35 (I)  \\
    &  arm-AFM & 5.98  &  6.76  & 1.98  & 1.96  & 1.98  & 118.1  & 123.6  & 1.3  & 0  & 4.52  & 3.91  & 1.55 (I)  \\
    
    \hline
    \hline 
  \end{tabular*}
\end{table*}

\begin{figure}[b]
\centering
  \includegraphics[width=8.5cm]{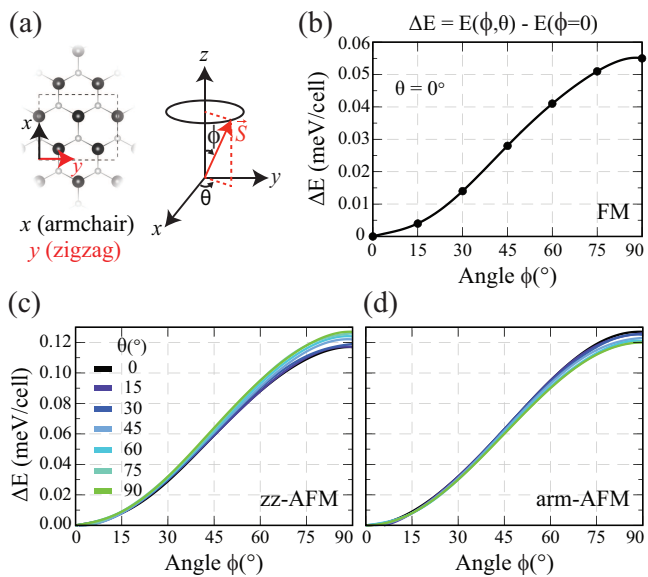}
  \caption{(a) Schematic illustration of the hexagonal lattice, and defined spherical coordinate system, in which $\theta$ is the angle between spin vector ($\vec{S}$) and x-axis and $\phi$ is the angle between $\vec{S}$ and z-axis. x and y lattice vectors label armchair and zigzag directions. Magnetic anisotropy energy (MAE) plots at spin vector directions for (b) AFM, (c) zz-AFM and (d) arm-AFM ordered 2D MnO structures.}
  \label{fig:2}
\end{figure}

Optimized structural parameteres of 2D MnO are obtained by taking each magnetic order into account. The calculated lattice parameters (\textit{a} and \textit{b}), and distances between Mn-O atoms (\textit{$d_{Mn-O_{1-3}}$}) are given in Table \ref{table1}. It is seen that while the  FM phase has a hexagonal lattice symmetry, the lattice favors a slightly distorted crystal structure when Mn atoms are antiferromagnetically (AFM) coupled. In addition, the distortion within the crystal lattice differs according to the prevailing AFM state. In zz-AFM, the crystal lattice shows  2.7 \% shrinkage and 1.2 \% enlargening along the armchair (\textit{a}) and zigzag directions (\textit{b}), respectively. On the other hand, in arm-AFM, \textit{a} and \textit{b} enlargens and narrows of about 0.2 \% and 1.9 \%, respectively. The variation in Mn-O-Mn angles with respect to the magnetic order are represented as $\alpha$ and $\beta$ within the Table \ref{table1}. The calculated bond length of 1.99 \AA{} between Mn-O atoms for the FM state, is found to differ in between 1.97-1.98 and 1.96-1.98 within the lattice structures of zz-AFM and arm-AFM order, respectively. Therefore, depending on the experimental procedure or synthesis techniques one can expect presence of internal strain inside the crystal and emergence of various magnetic phases in MnO single layers. 

As seen in Fig. \ref{fig:1}c, the magnetic phase transition from zz-AFM to arm-AFM, or vice versa, is estimated by applying a tensile strain of above 1.37 \% or a compressive strain of above approximately 1.34 \%, respectively. Therefore, strain engineering is useful tool to acquire controllable properties in 2D MnO. Furthermore, Bader charge analysis reveals that MnO lattice is formed by the ionically bonded Mn-O atoms, resulting from a charge donation of 1.3 $e^-$ from Mn to a O atom.

In order to determine the stability of magnetization against the thermal excitation, it is esential to investigate magnetocrystalline anisotropy in magnetic materials. For this purpose, non-collinear calculations are carried out for 2D MnO by rotating the spin vector through the angle $\phi$ and $\theta$, corresponding to the angle that the spin vector makes with the z- or x-axis (see Fig. \ref{fig:2}a). Here, magnetic anisotropy energy (MAE) per conventional cell is calculated by using the equation of $ \Delta E = E(\phi , \theta) - E( \phi = 0^{\circ})$, where $E(\phi , \theta)$ is the total energy of defined spin vector direction, and $E( \phi = 0^{\circ})$ is the energy of the easy axis. Fig. \ref{fig:2}(b-d) shows that the MAE per cell, the easy axis is perpendicular to the MnO basal plane ($\phi$ = 0$^{\circ}$) in all magnetic phases. The hard axis parallel to the basal plane lies along the zigzag ($\theta$ = 90$^{\circ}$) and armchair ($\theta$ = 0$^{\circ}$) directions for the structures consist of zz-AFM and arm-AFM orders, respectively. In the case of FM order, the total energy is independent from the varying $\theta$. 2D MnO displays small MAEs, which are calculated to be 0.055 meV for FM, and 0.127 meV for zz-AFM and arm-AFM ordered structures. The corresponding result reveals that single layer MnO is a soft antiferromagnetic material, where the direction of magnetization can be altered with small energies.  

\begin{figure}[t]
\centering
  \includegraphics[width=8.5cm]{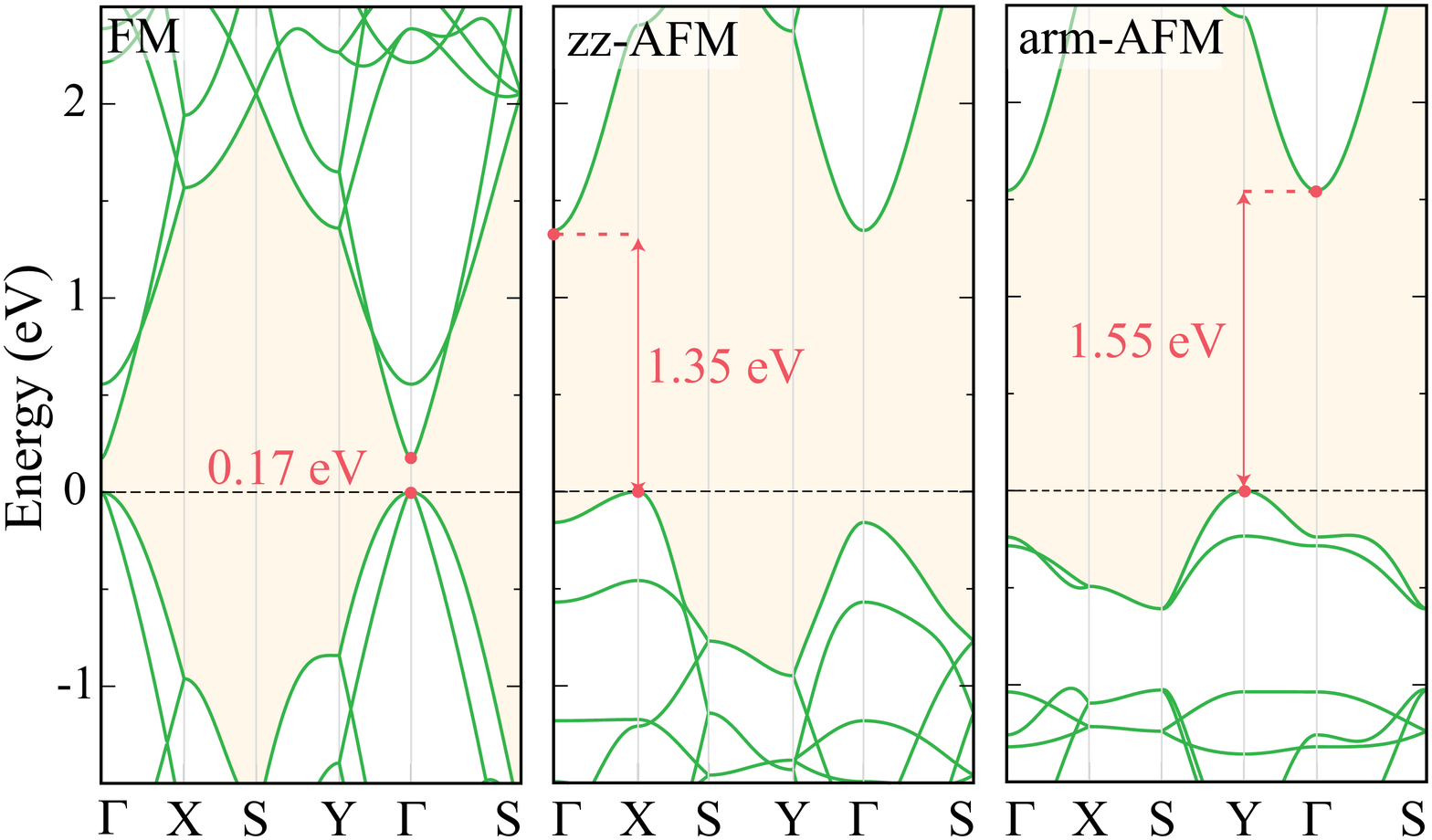}
  \caption{Electronic bandstructures calculated for FM, zz-AFM and arm-AFM ordered MnO single layer structures.}
  \label{fig:3}
\end{figure}

Electronic bandstructure calculations show that depending on the magnetic phase 2D MnO display quite different electronic characteristics. As shown in Fig. \ref{fig:3}, FM state causes a narrow direct band gap of 0.17 eV at the $\Gamma$ symmetry point. Conversely, AFM configurations causes an indirect band gap, whereas depending on its type, valence and conduction band edges (VBM and CBM) localize on distinctive zone points. In zz-AFM, the indirect band gap, corresponding to the energy difference between X and  $\Gamma$ point, is estimated as 1.35 eV. The arm-AFM order, where VBM and CBM appears at Y and $\Gamma$ point, respectively, yields comparably larger indirect band gap of 1.55 eV. Work functions ($\Phi$) are calculated to be 3.03, 3.72 and 3.91 for FM, zz-AFM and arm-AFM orders.

\subsection{Single Layer Nanoribbons of Manganese Oxide}

\subsubsection{Structural and magnetic properties}

\begin{figure}[t]
\centering
  \includegraphics[width=8.5cm]{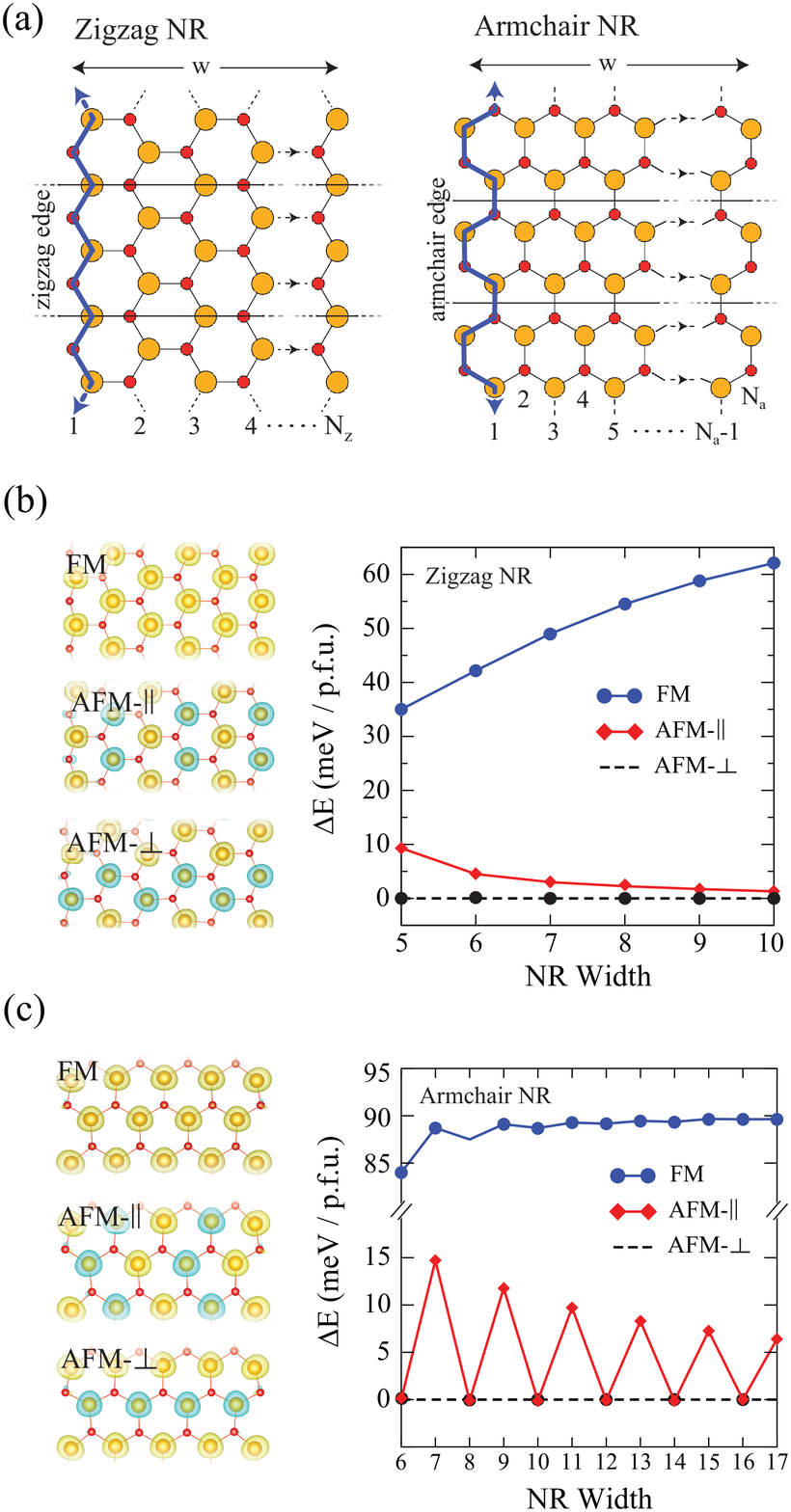}
  \caption{(a) Structural illustrations of MnO nanoribbons having zigzag and armchair edges. N$_z$ and N$_{a}$ corresponds to the number of zigzag chains and Mn-O dimer lines within the confined lattice, and indicate the ribbon width. The \textit{w} gives the distance between the opposite edge atoms. Plots in (b) and (c) show the variation in total energy difference per formula unit between FM, AFM-$\parallel$, and AFM-$\perp$ spin orders as a function of the ribbon width.}
  \label{fig:4}
\end{figure}

In this section, width dependent structural, magnetic and electronic properties of MnO nanoribbons are analyzed considering zigzag and armchair terminated edges. The width of the zigzag and armchair nanoribbons (ZNR and ANR) are determined by the number of zigzag chains (\textit{N$_z$}) and Mn-O dimer lines (\textit{N$_{a}$}) within the confined lattice direction, respectively. Depending on the width, ZNR and ANR structures are labelled as \textit{N$_z$}-ZNR and \textit{N$_{a}$}-ANR, where \textit{N$_z$} and \textit{N$_{a}$} ranges from 10 to 5, and 17 to 6, respectively. The representative illustration of each nanoribbon type is given in Fig. \ref{fig:4}a.

To understand the effect of confinement of charge carriers into the 1D on the magnetic properties, we investigate the energetics of FM and AFM configurations with respect to the varying ribbon width. Depending on the edge type, AFM orders with zigzag or armchair patterns can be perpendicular or parallel to the direction of the NR structures. Here, AFM orders in ribbons are labeled as AFM-$\parallel$ or AFM-$\perp$ with respect to the condition of being parallel or perpendicular to the repeating direction, respectively. Atomic structures of magnetic states and the variation in the total energy difference per formula unit ($\Delta$E) between them as a function of the ribbon width is represented in Fig. \ref{fig:4}b and c, for ZNR and ANR structures, respectively. As in the 2D structure of MnO, ribbon structures exhibit more favorable energetics in AFM spin configurations compared to FM ones. 

For ZNRs with larger surface area, the calculated total energies of AFM states are nearly identical, however resulting from the edge effects that become dominant in thinner structures, AFM-$\perp$ exhibits lower energetics compared to AFM-$\parallel$. While the energy difference between them is $\sim$1.3 meV in 10-ZNR, it increases to $\sim$9.3 meV in 5-ZNR. Even if ZNRs possess greater energy values in FM state, the $\Delta$E with respect to the magnetic ground state structure shows dramatic decrease with decreasing width. Moreover, owing to the dangling bonds at edges, the AFM-$\perp$ ordered ribbons display a small magnetic moment of 0.8 $\mu_B$ arising from the O atoms. On the contrary, the total magnetization in the AFM-$\parallel$ order oscillates from 1.2 to 8.8 $\mu_B$, depending on N$_z$ being even or odd number, respectively. In even case, the magnetism is purely from O atoms at edges, however, in odd case, additional magnetism appears as the number of Mn atoms that are aligned in the opposite direction is not equal. 

In ANRs, depending on whether \textit{N$_{a}$} is even or odd, we obtain an unsual energy trend resulting from the significant variation in the energy difference between AFM states. The reason is attributed to the glide or reflection symmetry between the opposite edges including even or odd number of dimer lines, respectively. Variation in total energy difference between magnetic states given in Fig \ref{fig:4}c reveals that the total energy difference between AFM orders becomes negligible if N$_{a}$ is even. However, when \textit{N$_{a}$} is odd, AFM-$\perp$ coupling displays lower energetics. The energy difference between AFM-$\parallel$ and AFM-$\perp$ tends to show gradual increase with a decreased width, such that, the $\Delta$E of $\sim$6.4 meV in 17-ANR is estimated to be $\sim$14.7 meV in 7-ANR. On the other hand, $\Delta$E between FM and ground state AFM distribution remains almost steady, except the slight drop in 6-ANR. Unlike the even-numbered ANR structures where the total magnetization completely vanishes, in odd-numbered structures, an additional Mn atom causes a magnetism of 5 $\mu_B$.

\subsubsection{Electronic properties}

\begin{figure}[t]
\centering
  \includegraphics[width=8.5cm]{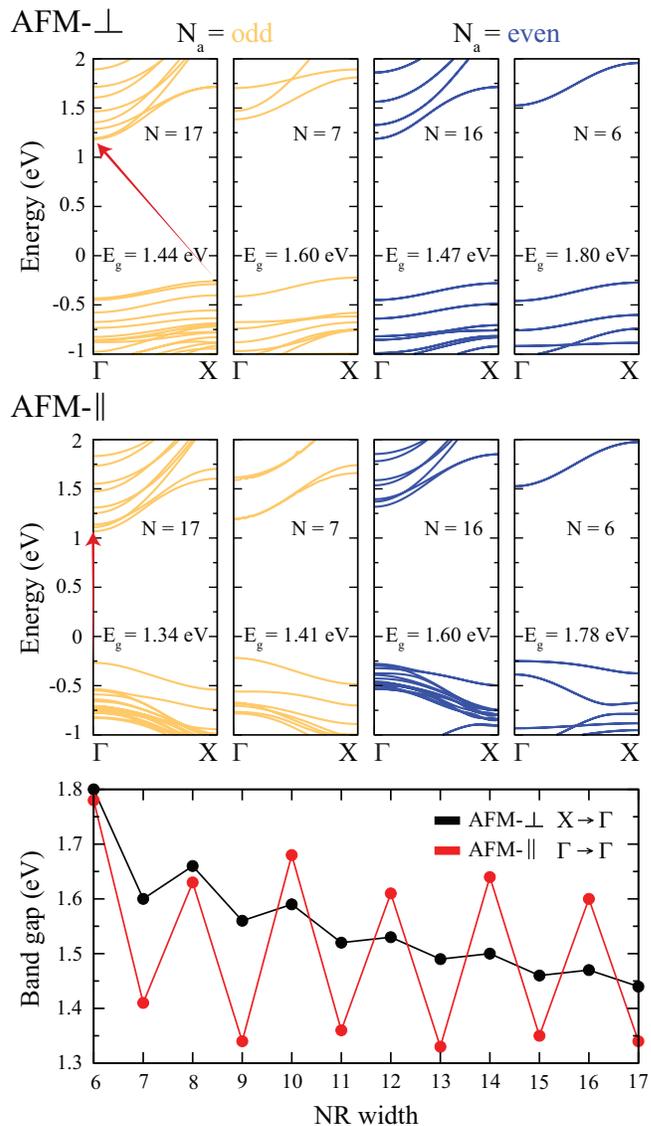}
  \caption{The calculated electronic band structures of 17-, 16-, 7-, and 6-ANR structures with respect to AFM-$\parallel$ and AFM-$\perp$ states. (b) The plot gives the variation in the band gap values as a function of the ribbon width.}
  \label{fig:5}
\end{figure}

Electronic properties of ZNR and ANR structures are investigated as a function of the ribbon width. Calculated electronic band structures show that ZNR structures exhibit metallic character due to the band crossing at the Fermi level caused by edge states. (see Fig. S1 in Supplementary Material) Since the magnetic ground state of ANRs depend on its width represented by \textit{N$_{a}$}, the investigation of electronic properties is carried out considering each type of AFM configuration. ANRs are predicted to be semiconductors whose band dispersion characteristics show significant dependency to the symmetry type in between the opposite edges, and the type of AFM state (see Fig. \ref{fig:5}). While ANRs with AFM-$\perp$ order are indirect band gap semiconductors, AFM-$\parallel$ order causes the formation of direct band gap at $\Gamma$ point. Apparently, the presence of Mn related magnetism in ANRs with odd N$_{a}$ cancels the degeneracy between minorty and majority states.   

In the case of AFM-$\perp$, the band gap values are inversely proportional to the width, and depending on whether the \textit{N$_{a}$} is even or odd, the change in the energy gap presents two distinct family behaviors. According to the plot in Fig \ref{fig:5}, which gives the evolution of band gap as a function of width, band gap in AFM-$\perp$ ordered structure widens with decreasing width by following a decent upward trend. As material gets thinner, odd-even based fluctuation in band gap value becomes more noticeable. The calculated band gap of 1.44 and 1.47 eV for 17-ANR and 16-ANR structures, enlarges to 1.60 and 1.80 eV for 7-ANR and 6-ANR structures, respectively. Furthermore, the charge density of lowermost conduction and topmost valence bands, labelled as CB and VB, are calculated through the $\Gamma-X$ path and illustrated in Fig. 2S for the specific ribbon widths. It is seen that CB is mostly dominated by the atomic orbitals of Mn-\textit{d} that localized at the edges. On the other hand, VB derived from the mixed Mn-\textit{d} and O-\textit{p} orbitals, which distribute through the inner lattice.

When AFM-$\parallel$ configuration is considered, the band gap values follow more complex behavior with the decreased width. As seen in Fig. 4, even-numbered ANRs induces significantly larger energy gap formation compared to those in odd-numbered ones. On the other hand, the gap size follows two distinct trends with respect to the condition of \textit{N$_{a}$} being equal to 4\textit{n}+1 or 4\textit{n}-1 (4\textit{n}+2 or 4\textit{n}-2) in odd (even) structures.  In here, \textit{n} is a positive integer. The VB and CB in AFM-$\parallel$ order consist of atomic orbitals mostly localized at edge atoms (see Fig. 3S). Due to the confinement of electron and hole states at edges, the gap size follows an almost steady trend down to the limit of the thinnest 6-ANR and 7-ANR structures, where it dramatically increases to 1.78 and 1.41 eV, respectively, resulting from the increased interaction between the opposite edge states. As it seen in Fig. 3S, the CB dominantly has Mn-\textit{d} atomic character, while the VB is Mn-\textit{d} and O-\textit{p} like. Moreover,the multiple family like behavior in AFM-$\parallel$ ordered ANR can be clarified with the spatial change in electron and hole states with varying ribbon width.

\section{Conclusions}\label{sec:conclusions}

In conclusion, this study is devoted to the first-principles investigation of structural, magnetic, and electronic properties of single layers of infinite 2D and nanoribbon structures of hexagonal MnO. Our total energy calculations revealed that while the ground state of MnO has a degenerate AFM character, there is also an excited FM state with relatively larger energy. As a result of their magnetic characteristics each phase also display distinct electronic properties. While AFM phases have large bandgap indirect semiconducting behavior, FM phase exhibits a direct narrow bandpag semiconducting behavior. The effect of one-dimensional confinement in MnO was investigated by means of zigzag and armchair edge terminated nanoribbon structures. Width-dependent total energy calculations revealed that AFM ordering persists down to the thinnest ribbon structures. In armchair nanoribbons having odd or even-numbered Mn-O dimer lines, the opposite edges display glide or reflection symmetry, which leads to a distinction in the comparable energetics of considered AFM magnetic orders. Unlike the zigzag edge formation that induces a metallicity, armchair-edged nanoribbons display semiconducting characteristics. The band gap is inversely proportional to the ribbon width and follows distinct trends originating from the variation in edge symmetry. This variation brings two and four different family behaviors when AFM-$\perp$ and AFM-$\parallel$ orders are considered, respectively. Furthermore, AFM-$\perp$ and AFM-$\parallel$ cause indirect and direct band gap formations in armchair nanoribbon structures. Our findings show that the low-dimensional structures of MnO with their edge and width dependent properties are promising materials to be used in nanoscale device applications. 

\section{Acknowledgments}
Computational resources were provided by TUBITAK ULAKBIM, High Performance and 
Grid Computing Center (TR-Grid e-Infrastructure). 

\bibliography{references}
\bibliographystyle{achemso}
\end{document}